\documentclass[%
prl,
 reprint,
 amsmath,amssymb,
 aps,preprintnumbers, longbibliography]{revtex4-1}

\usepackage[colorlinks=true,linkcolor=black, citecolor=black,
urlcolor=black]{hyperref}

\usepackage{multirow,graphics}
\usepackage{enumerate}
\usepackage{amstext}
\usepackage{amssymb}
\usepackage{amsmath}
\usepackage{graphicx}
\usepackage{color}
\usepackage{tikz}

\setcitestyle{square,numbers}

\begin{document}

\newcommand{\IM}{{\rm Im}\,}
\newcommand{\card}{\#}
\newcommand{\la}[1]{\label{#1}}
\newcommand{\eq}[1]{(\ref{#1})}
\newcommand{\figref}[1]{Fig. \ref{#1}}
\newcommand{\abs}[1]{\left|#1\right|}
\newcommand{\comD}[1]{{\color{red}#1\color{black}}}
\newcommand{\p}{\partial}
\newcommand{\Tr}{{\text{Tr}}}
\newcommand{\como}[1]{{\color[rgb]{0.0,0.1,0.9} {\bf \"O:} #1} }

\makeatletter
\newcommand{\subalign}[1]{%
  \vcenter{%
    \Let@ \restore@math@cr \default@tag
    \baselineskip\fontdimen10 \scriptfont\tw@
    \advance\baselineskip\fontdimen12 \scriptfont\tw@
    \lineskip\thr@@\fontdimen8 \scriptfont\thr@@
    \lineskiplimit\lineskip
    \ialign{\hfil$\m@th\scriptstyle##$&$\m@th\scriptstyle{}##$\crcr
      #1\crcr
    }%
  }
}
\makeatother

\newcommand{\mzvv}[2]{
  \zeta_{
    \subalign{
      &#1,\\
      &#2
    }
}
}

\newcommand{\mzvvv}[3]{
  \zeta_{
    \subalign{
      &#1,\\
      &#2,\\
      &#3
    }
}
  }

\makeatletter
     \@ifundefined{usebibtex}{\newcommand{\ifbibtexelse}[2]{#2}} {\newcommand{\ifbibtexelse}[2]{#1}}
\makeatother

\preprint{LPTENS--15/07, IPhT--T15/220}


\usetikzlibrary{decorations.pathmorphing}
\usetikzlibrary{decorations.markings}
\usetikzlibrary{intersections}
\usetikzlibrary{calc}

\tikzset{
photon/.style={decorate, decoration={snake}},
particle/.style={postaction={decorate},
    decoration={markings,mark=at position .5 with {\arrow{>}}}},
antiparticle/.style={postaction={decorate},
    decoration={markings,mark=at position .5 with {\arrow{<}}}},
gluon/.style={decorate, decoration={coil,amplitude=2pt, segment length=4pt},color=purple},
wilson/.style={color=blue, thick},
scalarZ/.style={postaction={decorate},decoration={markings, mark=at position .5 with{\arrow[scale=1]{stealth}}}},
scalarX/.style={postaction={decorate}, dashed, dash pattern = on 4pt off 2pt, dash phase = 2pt, decoration={markings, mark=at position .53 with{\arrow[scale=1]{stealth}}}},
scalarZw/.style={postaction={decorate},decoration={markings, mark=at position .75 with{\arrow[scale=1]{stealth}}}},
scalarXw/.style={postaction={decorate}, dashed, dash pattern = on 4pt off 2pt, dash phase = 2pt, decoration={markings, mark=at position .60 with{\arrow[scale=1]{stealth}}}}
}

 \newcommand{\doublewheelsmall}{
   \begin{minipage}[c]{1cm}
     
     \begin{center}
       \begin{tikzpicture}[scale=0.3]
         \foreach \m in {1,2} {
           \draw (0.75*\m,0) arc[radius = 0.75*\m,start angle = 0, end angle = 300] ;
           \draw[black, densely dotted] (300:0.78*\m) arc[radius = 0.75*\m,start angle = -60, end angle = 0];
         }
         \foreach \t in {1,2,...,5} {
           \draw (0,0) -- (60*\t:1.5);
         }
         \draw[black,densely dotted] (0,0) -- (0:1.5);
       \end{tikzpicture}
     \end{center}
   \end{minipage}
 }


\newcommand{\footnoteab}[2]{\ifbibtexelse{%
\footnotetext{#1}%
\footnotetext{#2}%
\cite{Note1,Note2}%
}{%
\newcommand{\textfootnotea}{#1}%
\newcommand{\textfootnoteab}{#2}%
\cite{thefootnotea,thefootnoteab}}}

\def\e{\epsilon}
     \def\bT{{\bf T}}
    \def\bQ{{\bf Q}}
    \def\wT{{\mathbb{T}}}
    \def\wQ{{\mathbb{Q}}}
    \def\ttQ{{\bar Q}}
    \def\tQ{{\tilde \bP}}
        \def\bP{{\bf P}}
        \def\dq{{\dot q}}
    \def\CF{{\cal F}}
    \def\cC{\CF}
    
     \def\l{\lambda}
\def\hbZ{{\widehat{ Z}}}
\def\bZ{{\resizebox{0.28cm}{0.33cm}{$\hspace{0.03cm}\check {\hspace{-0.03cm}\resizebox{0.14cm}{0.18cm}{$Z$}}$}}}

\title{New integrable non-gauge 4D QFTs  from strongly deformed planar N=4 SYM }

\author{\"Omer G\"urdo\u gan$^{a,b}$, Vladimir Kazakov$^{a,c}$}

\affiliation{%
             \(^{a}\) LPT, \'Ecole Normale Superieure, 24 rue Lhomond 75005 Paris, France \\
 \(^{b}\)Institut de Physique Th\'eorique, CEA Saclay, 91191 Gif-sur-Yvette Cedex, France\\
 \(^{c}\) Universit\'e Paris-VI, Place Jussieu, 75005 Paris, France
}

\begin{abstract}
  We introduce a family of new integrable quantum field theories in
  four dimensions by considering the \(\gamma\)-deformed
  \(\mathcal{N}=4\) SYM in the double scaling limit of large imaginary
  twists and small coupling. This limit discards the gauge fields and
  retains only certain Yukawa and scalar interactions with three
  arbitrary effective couplings. In the 't~Hooft limit, these 4D
  theories are integrable, and contain a wealth of conformal
  correlators such that the whole arsenal of AdS/CFT integrability
  remains applicable. As a special case of these models, we obtain a
  QFT of two complex scalars with a chiral, quartic interaction. The
  BMN vacuum anomalous dimension is dominated in each loop order by a
  single ``wheel'' graph, whose bulk represents an integrable
  ``fishnet'' graph. This explicitly demonstrates the all-loop
  integrability of gamma-deformed planar N=4 SYM, at least in our
  limit.  Using this feature and integrability results we   
  provided an explicit conjecture for the periods of double-wheel
  graphs with an arbitrary number of spokes in  terms of multiple
zeta values (MZV) of limited depth.
\end{abstract}

 \maketitle

\section{Introduction}

\({\cal N}=4\) SYM theory is an integrable CFT in the large \(N_c\)
limit \cite{Beisert:2010jr}, with an exactly and efficiently
calculable spectrum of anomalous dimensions via the quantum spectral
curve (QSC) formalism \cite{Gromov:2013pga,Gromov:2014caa}. It seemed
plausible that the gauge symmetry, and a large amount of supersymmetry
are necessary for its integrability. However, the twisted versions of
\({\cal N}=4\) SYM \cite{Leigh:1995ep,Frolov:2005dj,Beisert:2005if} loose the
supersymmetry partially, or even entirely, but remain integrable,
with the  twisted  QSC equations  \cite{Kazakov:2015efa,Gromov:2015dfa}.

This opens a window of opportunities for constructing new
non-supersymmetric QFTs, potentially more interesting for physical
applications. In this paper, we propose a special double-scaling (DS)
limit of the so-called \(\gamma\)-deformed twisted \({\cal N}=4\) SYM
which breaks the \(SU(4)\) R-symmetry down to \(U(1)^3\) subgroup,
combining large imaginary twists \(\gamma_j\to i\infty\) \footnote{A
  large \(q\) expansion was considered in \protect\cite{Gromov:2010dy}
  for \(\beta\)-deformation} and the weak coupling limit \(g^2\equiv
N_cg_{\mathrm{YM}}^2\to 0\), with three effective DS couplings
\(\xi_j^2=g^2e^{-i\gamma_j}\) kept fixed.  In this DS limit, the gauge
fields decouple and the theory reduces to a QFT of three complex
scalars and fermions interacting through quartic and ``chiral'' Yukawa
couplings. If only one coupling is turned on, the fermions and one
scalar decouple and we have the following bi-scalar action: \begin{align}
    \label{Lthree}
    {\cal L}_{\phi}= \frac{N_c}{2}\Tr\,\,
    \left(\p^\mu\phi^\dagger_1 \p_\mu\phi_1+\p^\mu\phi^\dagger_2 \p_\mu\phi_2+2\xi^2\,\phi_1^\dagger \phi_2^\dagger \phi_1\phi_2\right)\,.\hfill\,
  \end{align}
  These theories remain  integrable at any coupling and show a conformal behavior  in
  the large \(N_c\) limit.  This is a remarkable example of a
  4-dimensional QFT that is integrable in the 't~Hooft limit, even in the
  absence of supersymmetry and gauge symmetry!

  Strictly speaking, the conformality of \(\gamma\)-twisted \({\cal
    N}=4\) SYM is broken by the double-trace counterterms, in
  particular of the type \(\eta_{ij} \Tr\bigl(\phi_j^\dagger \phi_j
  \bigr)\Tr\bigl(\phi_j^\dagger\phi_j\bigr)\) and
  \(\tilde\eta_{ij}\Tr(\phi_i^\dagger \phi_j^\dagger)\Tr (\phi_i
  \phi_j) \) with the couplings \(\eta_{ij}\) and \(\tilde \eta_{ij}\) running with scale even at
  the leading large-\(N_c\) order~\cite{Tseytlin:1999ii,
    Dymarsky:2005uh, Fokken:2013aea, Jin:2013baa}). On the string
  theory side of the duality, this double trace anomaly was traced
  down to the tachyonic instability of the corresponding
  \(\gamma\)-deformed coset~\cite{Pomoni:2008de}. As it was recently
  pointed out in~\cite{Sieg:2016vap}, our new theories inevitably
  inherit the breakdown of conformality of the full gamma-deformed
  \(\mathcal{N}=4\) SYM. However, one can immediately see, e.g. by
  drawing the relevant one-loop Feynman diagrams (or simply adopting
  the results of~\cite{Jin:2013baa}) that \(\frac{\p
    \xi}{\p\log\mu}={\cal O} \left( N^{-2}\right)\) and all
  multi-point correlators which do not have length-\(2\) operators in
  the intermediate states are perfectly conformal in the 't~Hooft
  limit. Therefore, the double trace terms in the Lagrangian are
  irrelevant for our current work and we suppress them.
  \begin{figure}
    \centering
  \begin{minipage}[c]{0.45\linewidth}
\center  \includegraphics{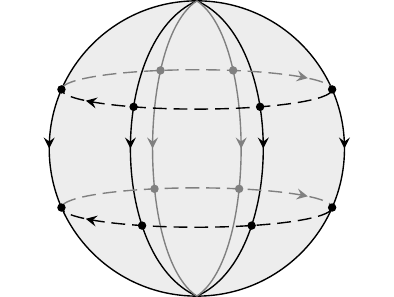}
\end{minipage}
\vspace{1cm}
\begin{minipage}{0.45\linewidth}
\center  \includegraphics{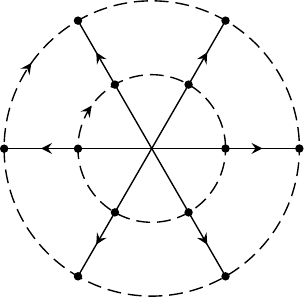}
\end{minipage}
\caption{The single possible Feynman graph contributing to the
  correlator of two the BMN vacuum operators in the model
  \eqref{Lthree} at a given order (given number \(M\) of wrappings):
  a) On the ``globe'' graph, \(\Tr \phi_1^{\dagger L}(x)\) and \(\Tr
  \phi_1^L(0) \) are placed at the north and south pole, respectively.
  \(L\) solid lines of \(\phi_1\) particles can be only crossed at the
  interaction vertices in one direction by \(M\) dashed lines of
  \(\phi_2\) particles.  b) The divergence of this graph is captured
  by the ``wheel'' graph, obtained by removing the vicinity of one of
  the poles.}
  \label{globe}
\end{figure}

Apart from the possibility of exact non-perturbative computations, the
integrability of this scalar QFT can serve for evaluating interesting
individual Feynman graphs, unavailable for other methods. The
anomalous dimension \(\gamma_\mathrm{vac}(L,z)\) of the BMN vacuum
operator \(\Tr \phi_1^L\) can be in principle computed exactly and the
dimension in each ``wrapping'' order is dominated by the period
(coefficient of the leading logarithm) of a single 4D Feynman graph of
a ``globe'' shape~\footnote{The idea that our limit can be dominated
  by ``wheel'' graphs arose in common discussion with M.~Wilhelm and
  Ch.Sieg}, see Fig.\ref{globe}.  Hence it can be computed from the
corresponding (yet to be found) DS limit of twisted QSC equations
\cite{Kazakov:2015efa} for any number of wrappings \(M\) (dashed
\(\phi_2\) lines on the Fig.\ref{globe}).  For \(M=1\), the result was
known from a direct graph computation
\cite{Broadhurst:1985vq,Fokken:2014soa} and for \(M=2\), from
AdS/CFT Y-system/TBA computations \cite{Ahn:2011xq} in terms of
infinite sums and integrals. We will present here the latter result
for \(M=2\) and any \(L\) in an explicit form containing only multiple
zeta values (MZV) of limited depth. This result allows us to extract
the periods of double-wheel graphs at an arbitrary loop order.

\begin{figure}
  \centering
  \includegraphics{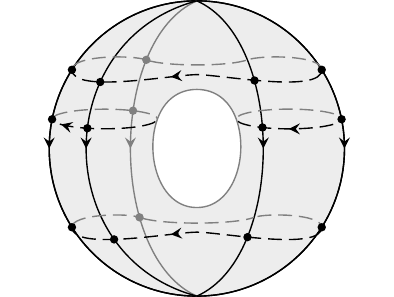}
  \caption{A graph of torus topology contributing to the  two-point function of two BMN vacuum operators}
  \label{fig:torus}
\end{figure}

  At finite \(N_c\), the theory \eqref{Lthree} (as well as the more
  general \eqref{eq:LDS}) is not conformal anymore.  The \(1/N^2_c\)
  corrections to the dimension of the operator \(\Tr\,\phi_1^L\) will
  be dominated by graphs depicted in Fig.\ref{fig:torus}. Notice that
  at each loop level we have to sum over distributions of
  \(L=L_r+L_l\) lines between right and left sides of the torus and
  the appearance of the singular cases \(L_r=2\) or \(L_l=2\), which
  break conformality, is unavoidable. Hence the addition of the
  aforementioned counterterms is also imposed to cancel the UV
  divergences.

\section{ Chiral boson QFT from strongly  \(\gamma\)-deformed N=4 SYM}

The lagrangian of \(\gamma\)-deformed \({\cal\ N}=4\) SYM reads  (see e.g.\cite{Fokken:2013aea})
\begin{multline}\label{eq:L}
\!\!\!\!\!\!\!  {\cal L}=N_c\Tr\biggl[
  -\frac{1}{4} F_{\mu\nu}F^{\mu\nu}
  -\frac{1}{2}D^\mu\phi^\dagger_iD_\mu\phi^i
  +i\bar\psi^{\dot\alpha}_{ A}D^\alpha_{\dot\alpha}\psi^A_{\alpha }\biggr]
+{\cal L}_{\rm int}
\end{multline}
where \(i=1,2,3\)\,\, \(A =1,2,3,4\),
\(D^{\alpha}_{\dot\alpha}= D_\mu
(\tilde\sigma^{\mu})^\alpha_{\dot\alpha}\) with
\((\tilde\sigma^{\mu})^\alpha_{\dot\alpha}=(-i\sigma_2,i\sigma_3,
\mathbb{I},-i\sigma_1)^\alpha_{\dot\alpha}\,\) and 
 \begin{equation*}
   \begin{aligned}[y]
     &\mathcal{L}_{\rm int} =N_cg\,\,\Tr\bigl[\frac{g}{4} \{\phi^\dagger_i,\phi^i\}
     \{\phi^\dagger_j,\phi^j\}-g\,e^{-i\epsilon^{ijk}\gamma_k}
     \phi^\dagger_i\phi^\dagger_j\phi^i\phi^j\\
     &-e^{-\frac{i}{2}\gamma^-_{j}}\bar\psi^{}_{ j}\phi^j\bar\psi_{ 4}
     +e^{+\frac{i}{2}\gamma^-_{j}}\bar\psi^{}_{ 4}\phi^j\bar\psi_{ j}
     + i\epsilon_{ijk} e^{\frac{i}{2} \epsilon_{jkm} \gamma^+_m} \psi^k \phi^i \psi^{ j}\\
     &-e^{+\frac{i}{2}\gamma^-_{j}}\psi^{}_{ 4}\phi^\dagger_j\psi_{
       j}
     +e^{-\frac{i}{2}\gamma^-_{j}}\psi^{}_{j}\phi^\dagger_j\psi_{
       4}
     + i\epsilon^{ijk} e^{\frac{i}{2} \epsilon_{jkm} \gamma^+_m} \bar\psi_{ k} \phi^\dagger_i \bar\psi\bigr]\hfill\, .
   \end{aligned}
 \end{equation*}
where the summation is assumed w.r.t. doubly and triply repeating indices. We suppress the  Lorentz indices of fermions, assuming the contractions
  \(\psi_i^\alpha \psi_{j,\alpha}\) 
 and  \(\bar\psi_{i,\dot\alpha} \bar\psi_j^{\dot\alpha}\). We also use the notations
 \begin{equation*}
\gamma_1^{\pm}=-\frac{\gamma_3\pm\gamma_2}{2},\quad\!\!
\gamma_2^{\pm}=-\frac{\gamma_1\pm\gamma_3}{2},\quad\!\!
\gamma_3^{\pm}=-\frac{\gamma_2\pm\gamma_1}{2}\, .
\end{equation*}
The parameters of the $\gamma$-deformation 
{$q_j=e^{-\frac{i}{2}\gamma_{j}}$ $j=1,2,3$} are related related to
the 3 Cartan subgroups of \(SU(4) \cong SO(6)\).

 We propose the following double scaling limit of the \(\gamma\)-deformed Lagrangian (\ref{eq:L}): 
 \begin{eqnarray*}&& q_3 \sim q_2 \sim q_1 \to\infty, \qquad g\to 0,\\
   && \xi_1:= q_1 g,\quad \xi_2:= q_2 g, \quad \xi_3:= q_3 g \quad
   \text{fixed},\end{eqnarray*} where the large \(q_i\) limit
 corresponds to sending \(\gamma_i\to i \infty\). Notice that, as a
 consequence of imaginary \(\gamma_j\), the action is not real
 anymore, and hence the unitarity of the theory is violated
 ~\footnote{We thank K.~Zarembo for the discussion and very helpful
   comments on this issue}. But the resulting QFT looks still very
 interesting, especially due to its integrability in the 't~Hooft
 limit. It is easy to see that in this DS limit of
 \(\mathcal{L}_{int}\) only some Yukawa and 4-scalar interactions
 survive and we arrive at the following integrable QFT of complex
 scalars and fermions (but no gauge fields):
\begin{multline}\label{eq:LDS}
  {\cal L}_{\phi\psi}=N_c\Tr\bigl[-\frac{1}{2}\p^\mu\phi^\dagger_i\p_\mu\phi^i
  +i\bar\psi^{\dot\alpha}_{ A}(\tilde\sigma^{\mu})^\alpha_{\dot\alpha}\p_\mu \psi^A_{\alpha }\bigr]
+{\cal L}_{\rm int}
\end{multline}
where
\begin{equation*}
   \begin{aligned}[y]
     \mathcal{L}_{\rm int} =N_c&
\,\Tr\bigl[\xi_1^2\,\phi_2^\dagger \phi_3^\dagger \phi_2\phi_3+\xi_2^2\,\phi_3^\dagger \phi_1^\dagger \phi_3\phi_1+\xi_3^2\,\phi_1^\dagger \phi_2^\dagger \phi_1\phi_2\hfill\\
&
\begin{aligned}[t]
 +&i\sqrt{\xi_2\xi_3}(\psi^3 \phi^1 \psi^{ 2}+ \bar\psi_{ 3} \phi^\dagger_1 \bar\psi_2 )\\
 +&i\sqrt{\xi_1\xi_3}(\psi^1 \phi^2 \psi^{ 3}+ \bar\psi_{ 1} \phi^\dagger_2 \bar\psi_3 )\\
 +&i\sqrt{\xi_1\xi_2}(\psi^2 \phi^3 \psi^{ 1}+ \bar\psi_{ 2} \phi^\dagger_3 \bar\psi_1 )\,\bigr].
\end{aligned}
   \end{aligned}
 \end{equation*}
In particular, we can send here \(\xi_1\to 0,\,\,\xi_2\to 0\). Then the fermions and one of the scalars decouple and we get a simpler bi-scalar action \eqref{Lthree}
where we denoted \(\xi\equiv\xi_3=q_3 g\).
We will provide some signs and consequences of  its integrability in the following sections.

\section{dimension of the BMN vacuum and ``wheel'' graphs}

Figure \ref{globe} demonstrates our statement in the introduction: in
our theory \eqref{Lthree} at \(N_c=\infty\) the correlator \(\left<\Tr
  \phi_1^{\dagger L}(x)\Tr\phi_1^L(0)\right>\) is dominated by
globe-like graphs. The point-splitting procedure gives anomalous
dimension of this operator at a given wrapping order \(M\) in terms of
the periods of wheel graphs with \(L\) spokes and \(M\) frames. It was
argued in \cite{Panzer:2015ida} that for $L=3$ and for any $M$, all
such periods can be expressed in terms of multiple zeta
values. However for \(L \geq 3\) and \(M\geq 2\), such a property is
not immediately obvious in existing approaches to loop integrals
\footnote{We are grateful to E. Panzer for his comments on this class
  of integrals.}. On the other hand, these dimensions can in principle
be computed using exact integrability tools of \({\cal\ N}=4\) SYM
applied to this model, such as the twisted QSC of
\cite{Kazakov:2015efa,Kazakov:2015efa}.

The anomalous dimension for this BMN vacuum operator, for arbitrary
\(\gamma\)-twist, up to two wrappings was obtained in
\cite{Ahn:2011xq} using the early Y-system/TBA version of spectral
equations
\cite{Gromov:2009tv,Bombardelli:2009ns,Gromov:2009bc,Arutyunov:2009ur}
(see also the QSC computation at one wrapping
\cite{Kazakov:2015efa}). This state is characterized only by two twist
parameters \(q_3,q_2\) which are denoted in \cite{Ahn:2011xq} as
\(q=(q_3q_2)^{L/2}\) and \(\dq=(q_3/q_2)^{L/2}\), so that in the DS
limit \(q\gg \dq \gg 1\) and \(z^2=\xi^{2L}=g^{2L}q\dq\) is fixed.
In our DS limit, the result of \cite{Ahn:2011xq} (see eqs. (5.5-7)
there) reduces to
\begin{equation}
  \label{gammaVAC}
  \gamma_\mathrm{vac}(L)
  =
  \gamma^{(1)}_\mathrm{vac}(L)\,z^2
  +
  \gamma^{(2)}_\mathrm{vac}(L)\,z^4
  +
  \mathcal{O}(z^6),
\end{equation} 
where \(\gamma^{(1)}_\mathrm{vac}(L) =
-2\,\binom{2L-2}{L-1}\,\zeta_{2L-3} \) and
\(\gamma^{(2)}_\mathrm{vac}(L)\) was computed in \cite{Ahn:2011xq} in
terms of double integrals and infinite double sums over digamma
functions.  We have computed \(\gamma^{(2)}_\mathrm{vac}(L)\) explicitly for any $L\geq 3$,
and present our result in the appendix \footnote{The singularity at \(L=2\)
  in this formula signals the breakdown of the conformal invariance
  \protect\cite{Fokken:2013aea} but this \(L=2\) state does not appear
  in most of the multi-point correlators in planar limit.}.

Here we discuss some consequences of this result and find evidence for
that it really computes the wheel graphs corresponding to our
bi-scalar action. Firstly, the $\mathcal{O}(z^2)$ term, corresponding
to a single wrapping, has been computed from standard Feynman diagram
technique in \cite{Fokken:2014soa} for arbitrary twists. In our DS limit,
this coefficient receives leading-order contributions only from the
bosons, cf. equation (3.4) of \cite{Fokken:2014soa}, and is given by the
periods of wheel integrals which were computed in
\cite{Broadhurst:1985vq}.

This argument for the existence of a single graph holds also at higher
wrappings and for two wrappings, i.e. \(\mathcal{O}(z^4) \), the anomalous
dimension is given in terms of the periods of double-wheel graphs:
  \begin{equation}
    \label{eq:gammatwowrap}
    \gamma_{\mathrm{vac}}^{(2)}(L)
    =
    2^{1-4L}\,
    \mathcal{P}\,\biggl(\doublewheelsmall\biggr)\, .
  \end{equation}
In support of (\ref{eq:gammatwowrap}), for \(L=3\) we have from
\cite{Gromov:2009tv} or from our general formula \eqref{eq:result}
\begin{equation*}
    \gamma^{(2)}_{\mathrm{vac}}(3)
    =
    2^{-11}\,\biggl(
    \frac{189}{2}\zeta_7 - 72 \zeta_3^2
    \biggr)\, .  
\end{equation*}
This value is reproduced by the period of ``double-wheel'' graph computed
explicitly in \cite{Panzer:2013cha} (see eq. (5.12) there).

Setting \(L=4\) in our general formula \eqref{eq:result} and using
various identities between the MZVs of weight 11 we find for example the following new result:
\begin{multline}     
  \gamma^{(2)}_{\mathrm{vac}}(4) =
  2^{-11}\,\bigl[
  \frac{309}{4}\zeta_{11}
  +4\zeta_{3,8}
  +5\zeta_{5,6}
  -\zeta_{6,5}\\
  \hfill
  +10\zeta_{8,3}
  -2\zeta_{3,3,5}
  +10(\zeta_{3,5,3}+\zeta_{5,3,3})
  -50\, \zeta_{5}^2
  \bigr]\, .
\end{multline}

\section{Integrability and QSC}

Leaving the derivation of the DS limit for these models from the
general twisted QSC (TQSC) equations of \cite{Kazakov:2015efa} for
future works in progress, we discuss here the underlying spin chain
picture. Taking as example the operators of Figure~\ref{globe}, we
notice that the computation of dominant wheel-type graphs can be
viewed as a certain hamiltonian evolution in the radial direction: It
implies the introduction of a transfer matrix of an equivalent quantum
spin chain adding one more dashed line on Fig.\ref{globe}. This
transfer matrix has the form \footnote{We are grateful to K.~Zarembo
  for sharing with us his computations of single wrapping case using
  similar techniques}:
   \begin{equation}\label{spinchainT}
\hat T=\xi^{2L}\prod_{l=1}^{L}\frac{1}{(x_{l+1}-x_{l})^2}\prod_{l=1}^{L}\Delta_{x_l
}^{-1},\quad (x_{L+1}\equiv x_1) \, ,
\end{equation} where \(x_1,\dots, x_L\) represent the 4D coordinates of interaction
vertices along the dashed line, and \(\Delta_{x_l}\) are the corresponding 4D Laplacians.  The first factor corresponds to adding a scalar propagator in
the angular direction on the graph, and \(\Delta_{x_l}^{-1}\)   -- adding a propagator in
the radial direction. This   spin chain  with conformal   \(SL(2,2)\)  symmetry is shown to be {\it\ integrable} in \cite{Zamolodchikov1980a}, due to the  star-triangle relations for ``fishnet'' graphs. This provides us with the proof, from first principals, of  all-loop integrability of ($\gamma$-deformed) \(\mathcal{N}=4\) SYM, at least in our rather non-trivial DS limit, without appeal to the still mysterious AdS/CFT duality. It also   should be the best starting point for   assembling more complicated physical quantities, such as correlators and \(1/N_c^2\) corrections. One can easily notice that the regular square lattice of Feynman integrals (so called ``fishnet''
graph)  is the most essential ``bulk'' part of any Feynman diagram for any physical quantity at sufficiently large loop order. It is also interesting to note that several physical quantities have only a single graph contribution at any nonvanishing order of perturbation theory, apart from some degenerate examples. This was already observed in the zero-dimensional analogue of this bi-scalar model  proposed in~\cite{Kostov:1996bs}.

 \section{Discussion}

 Our work shows that integrability in 4D in 't~Hooft limit can exist
 not only in the absence of supersymmetry but even without gauge
 symmetry. In this discussion we comment on open problems and some
 properties of our new models yet to be precised. An integrable model
 of interacting fermions and complex scalar fields, similar to
 \eqref{eq:LDS} but with sextic scalar interactions, exists also in 3D
 and can be obtained by a similar large twist/small coupling DS limit
 from the $\gamma$-deformed ABJM model~\cite{CGKtbp}.  All these
 models are chiral and have complex actions and are therefore
 non-unitary. Nevertheless, they may be rather interesting for various
 particle physics and statistical mechanics applications. They could
 be considered as a non-trivial zero-order approximation to more
 familiar theories. For example, as was discussed above, these
 theories have a non-zero beta function and generate a mass scale
 already at the leading large \(N_c\) order, a phenomenon worth
 studying by the use of integrability. Since, these theories
 nevertheless show conformal behavior for a multitude  of the correlators in
 the 't~Hooft limit, they can be used to verify numerous existing
 hypothesis concerning CFTs at dimensions \(D>2\).

 It is worth noticing that these QFTs obey a kind of chirality
 property on their planar Feynman graphs, suggesting a ``string
 worldsheet'' picture: if we associate the ``worldsheet time''
 \(\tau\) to solid lines on Figures \ref{globe} and \ref{fig:torus},
 then the dashed lines, associated with space direction \(\sigma\)
 have a particular orientation in, say, clockwise direction, i.e. a
 particular ``worldsheet'' chirality.  In the opposite double-scaling
 limit of small twists we would reproduce a similar QFT with the
 opposite chirality on planar graphs. Note that the transfer matrix
 \eqref{spinchainT} corresponds to the choice of radial ``worldsheet
 time'' on this planar graph.  However, the string dual of these
 theories is not evident: the original AdS/CFT interpretation is not
 directly applicable due to the weak coupling limit of the twisted
 \({\cal N}=4\) SYM, with the AdS radius collapsing to zero.

 For more complicated physical quantities, such as the correlators of
 other operators mixing both scalar fields, three point functions
 (structure constants), etc., the dominating Feynman graphs will be
 more complicated.  However their classification, due to chirality, is
 significantly simpler than in the general \({\cal N}=4\) SYM
 case~\cite{CGKtbp}.  Remarkably, many multi-point correlators are
 given by a single diagram at each order of perturbation theory for a
 given configuration of operators. This should ease the computation of
 these quantities directly by using the integrable transfer matrix
 \eqref{spinchainT} as the principal building block.

 Furthermore, integrability can help to compute exactly some classes
 of complicated {\it individual} Feynman graphs at any loop order,
 which is so far unavailable from the other methods.  We demonstrated
 this here on the example of BMN vacuum and the computation of wheel
 graphs. In a forthcoming work~\cite{CGKtbp} we analyse a slightly
 larger class of multi-magnon operators of the type \({\rm
   tr}(\phi_{j_1}\phi_{j_2}\dots\phi_{j_L}) \), with \(j_k=1,2\). The
 corelators of these are given by graphs in which the parallels in
 Figure \ref{globe} are replaced by non-intersecting spirals that
 connect the two poles. For multi-point correlators, the classification
 of Feynman graphs needs additional efforts.

 Finally, we should stress that our observation of the explicit
 integrability of correlators and Feynman integrals does not mean that
 the answers are immediately available for any interesting
 quantity. As the case of wheel graphs shows, a considerable effort is
 required to obtain for the simplest, although remarkable,
 results. The (yet to be established) explicit double-scaling form of
 the QSC equations should allow for more efficient analytic and
 numerical studies of these curious theories, both perturbatively and
 non-perturbatively.

  
\appendix*            
\setcounter{equation}{0}

\section{Appendix: Explicit formula for the double-wheel graphs}
\label{appendix}
In this appendix we provide our result for the double-wrapping
contribution to \(\gamma_{\mathrm{vac}}\) in (\ref{gammaVAC}) which are
given by the periods of double-wheel graphs.

\begin{widetext}
  \begin{multline}\label{eq:result}
    \gamma^{(2)}(L)=\frac{2^{3}}{\Gamma^2(L)}\Biggl\{
    -\sum _{j_1=0}^{L-1} \frac{ (2 L-2)!}{L-j_1}  \binom{L+j_1-1}{j_1}
    \zeta_{2 L-3}\mzvv{L+j_1-2}{L-j_1}
    +\frac{\Gamma^2(L)}{8}{4L-2 \choose 2L -1} \zeta_{4L-5}
    \hfill
    \\
    +\sum_{\substack{j_1, j_2 > 0\\j_1+j_2< 2L-3}} 
    \frac{\Gamma(2L-j_1-j_2-1) \Gamma (L+j_1) \Gamma(L+j_2)}{\Gamma
      (j_1+1) \Gamma (j_2+1) \Gamma (L-j_1) \Gamma (L-j_2)} \Biggl[
    \sum_{k=1}^{L+j_1-2} \Biggl(
    \binom{L+k+j_2-4}{k-1}(2\mzvvv{L+j_1-k-1}{L+j_2+k-3}{2L-j_1-j_2-1}+\mzvv{L+j_1-k-1}{3L-j_1+k-4})\\
    -\binom{L+k+j_2-2}{k}(2\mzvvv{L+j_1-k-1}{L+j_2+k-2}{2L-j_1-j_2-2}-2\mzvvv{L+j_1-k-1}{L+j_2+k-1}{2L-j_1-j_2-3})    
    \Biggr)
    +2\mzvv{2L+j_1+j_2-3}{2L-j_1-j_2-2}
    +2\mzvvv{2L+j_2-1}{L+j_1-1}{2L-j_1-j_2-3}
    \mzvv{L+j_1-2}{3L-j_1-3}
    -\mzvv{2L+j_1+j_2-2}{2L-j_1-j_2-3}
    \Biggr]\\
    +\frac{\Gamma^2(2L-1)}{2\Gamma^2(L)}
    \Biggl[
    \sum _{k=1}^{2 L-4} \Biggl(
    {2L+k-4 \choose k}
    \bigl(
    3 \mzvv{2L-k-3}{2L+k-2}
    +2 \mzvvv{2L-k-3}{2L+k-4}{2}
    \bigr)
    -2\Biggl( {2L+k-4 \choose k-1} + {2L+k-4 \choose k+1}\Biggr)\mzvvv{2L-k-3}{2L+k-3}{1}
    \Biggr)\\
    +2\mzvvv{2L-3}{2L-4}{2}
    -4(L-3)\mzvvv{2L-3}{2L-3}{1}
    +\mzvv{2L-4}{2L-1}
    -4\mzvv{2L-2}{2L-3}
    +4\mzvv{4L-6}{1}
    +2\mzvvv{3L-4}{2L-2}{1}
    +\frac{2L+1}{L}\zeta_{2L-3}\zeta_{2L-2}
    -5\zeta_{4L-5}
    -\zeta_{2L-3}^2
    \Biggr]
\Biggr\}\, ,
  \end{multline}
\end{widetext}
where \(j_1\) \(j_2\) in the double sum are non-negative and take values
such that \(j_1+j_2 <2L-3\). Furthermore, we have used, for visual convenience, 
the following (rather unconventional) notation for the MZVs:
\begin{equation}
  \label{eq:MZVnot}
 \mzvvv{w_1}{\raisebox{3pt}{\parbox{1em}{\centering\scalebox{0.5}{\vdots}}}}{w_r} := \sum_{1<n_1<\dotsm<n_r}^{\infty}\prod_{i=1}^r\frac{1}{n_i^{w_i}}.
\end{equation}
Although our result for $\gamma^{(2)}(L)$ is sizeable, it has some noteworthy
features. Its transcendentality weight is almost uniform and equal to
$4L-5$ with the exception of a weight-$(4L-6)$ piece, which amounts to
\begin{equation*}
  \label{eq:smallweight}
  \gamma^{(2)}(L)\bigr|_{\text{weight}-(4L-6)}
  = -2^{2-4L}\frac{\Gamma^2(2L-1)}{\Gamma^4(L)}
  \zeta_{2L-3}^2\, .
\end{equation*}
Moreover it is given in terms of MZVs of depth up to three,
irrespective of the weight. We provide an expression for $\gamma^{(2)}(L)$ in a
Mathematica-compatible fortmat in the attached file \verb+gamma2L.m+ .

\paragraph{Note added:} On the day of the submission of the second
version of our paper, the letter arXiv:1602.05817 on the
non-conformality of our model has appeared. We find no contradiction
with our arguments and precise our point of view in the updated
version.

\begin{acknowledgments}
\section*{Acknowledgments}
\label{sec:acknowledgments}

We thank D.Bernard, E.Brezin, J.Iliopoulos, J.Fokken, G.Korchemsky,
L.Hollo, E.Panzer, Ch.Sieg, G.Sizov, S. van Tongeren, D.Volin,
P.Wiegmann, M.Wilhelm and especially J.Caetano and K.Zarembo for
discussions.  The work of \"O.G. and V.K. was supported by the People
Programme (Marie Curie Actions) of the European Union's Seventh
Framework Programme FP7/2007-2013/ under REA Grant Agreement No 317089
(GATIS), by the European Research Council (Programme ``Ideas''
ERC-2012-AdG 320769 AdS-CFT-solvable), from the ANR grant StrongInt
(BLANC- SIMI- 4-2011). V.K. is grateful to Humboldt University
(Berlin) for the financial support of this work in the framework of
the ``Kosmos'' program.

\end{acknowledgments}

\appendix

\bibliography{biblio}

\begin{thebibliography}{33}%
\makeatletter
\providecommand \@ifxundefined [1]{%
 \@ifx{#1\undefined}
}%
\providecommand \@ifnum [1]{%
 \ifnum #1\expandafter \@firstoftwo
 \else \expandafter \@secondoftwo
 \fi
}%
\providecommand \@ifx [1]{%
 \ifx #1\expandafter \@firstoftwo
 \else \expandafter \@secondoftwo
 \fi
}%
\providecommand \natexlab [1]{#1}%
\providecommand \enquote  [1]{``#1''}%
\providecommand \bibnamefont  [1]{#1}%
\providecommand \bibfnamefont [1]{#1}%
\providecommand \citenamefont [1]{#1}%
\providecommand \href@noop [0]{\@secondoftwo}%
\providecommand \href [0]{\begingroup \@sanitize@url \@href}%
\providecommand \@href[1]{\@@startlink{#1}\@@href}%
\providecommand \@@href[1]{\endgroup#1\@@endlink}%
\providecommand \@sanitize@url [0]{\catcode `\\12\catcode `\$12\catcode
  `\&12\catcode `\#12\catcode `\^12\catcode `\_12\catcode `\%12\relax}%
\providecommand \@@startlink[1]{}%
\providecommand \@@endlink[0]{}%
\providecommand \url  [0]{\begingroup\@sanitize@url \@url }%
\providecommand \@url [1]{\endgroup\@href {#1}{\urlprefix }}%
\providecommand \urlprefix  [0]{URL }%
\providecommand \Eprint [0]{\href }%
\providecommand \doibase [0]{http://dx.doi.org/}%
\providecommand \selectlanguage [0]{\@gobble}%
\providecommand \bibinfo  [0]{\@secondoftwo}%
\providecommand \bibfield  [0]{\@secondoftwo}%
\providecommand \translation [1]{[#1]}%
\providecommand \BibitemOpen [0]{}%
\providecommand \bibitemStop [0]{}%
\providecommand \bibitemNoStop [0]{.\EOS\space}%
\providecommand \EOS [0]{\spacefactor3000\relax}%
\providecommand \BibitemShut  [1]{\csname bibitem#1\endcsname}%
\let\auto@bib@innerbib\@empty
\bibitem [{\citenamefont {Beisert}\ \emph {et~al.}(2012)\citenamefont {Beisert}
  \emph {et~al.}}]{Beisert:2010jr}%
  \BibitemOpen
  \bibfield  {author} {\bibinfo {author} {\bibfnamefont {Niklas}\ \bibnamefont
  {Beisert}} \emph {et~al.},\ }\bibfield  {title} {\enquote {\bibinfo {title}
  {{Review of AdS/CFT Integrability: An Overview}},}\ }\href {\doibase
  10.1007/s11005-011-0529-2} {\bibfield  {journal} {\bibinfo  {journal} {Lett.
  Math. Phys.}\ }\textbf {\bibinfo {volume} {99}},\ \bibinfo {pages} {3--32}
  (\bibinfo {year} {2012})},\ \Eprint {http://arxiv.org/abs/1012.3982}
  {arXiv:1012.3982 [hep-th]} \BibitemShut {NoStop}%
\bibitem [{\citenamefont {Gromov}\ \emph {et~al.}(2014)\citenamefont {Gromov},
  \citenamefont {Kazakov}, \citenamefont {Leurent},\ and\ \citenamefont
  {Volin}}]{Gromov:2013pga}%
  \BibitemOpen
  \bibfield  {author} {\bibinfo {author} {\bibfnamefont {Nikolay}\ \bibnamefont
  {Gromov}}, \bibinfo {author} {\bibfnamefont {Vladimir}\ \bibnamefont
  {Kazakov}}, \bibinfo {author} {\bibfnamefont {Sebastien}\ \bibnamefont
  {Leurent}}, \ and\ \bibinfo {author} {\bibfnamefont {Dmytro}\ \bibnamefont
  {Volin}},\ }\bibfield  {title} {\enquote {\bibinfo {title} {{Quantum Spectral
  Curve for Planar $\mathcal{N} =$ Super-Yang-Mills Theory}},}\ }\href
  {\doibase 10.1103/PhysRevLett.112.011602} {\bibfield  {journal} {\bibinfo
  {journal} {Phys. Rev. Lett.}\ }\textbf {\bibinfo {volume} {112}},\ \bibinfo
  {pages} {011602} (\bibinfo {year} {2014})},\ \Eprint
  {http://arxiv.org/abs/1305.1939} {arXiv:1305.1939 [hep-th]} \BibitemShut
  {NoStop}%
\bibitem [{\citenamefont {Gromov}\ \emph {et~al.}(2015)\citenamefont {Gromov},
  \citenamefont {Kazakov}, \citenamefont {Leurent},\ and\ \citenamefont
  {Volin}}]{Gromov:2014caa}%
  \BibitemOpen
  \bibfield  {author} {\bibinfo {author} {\bibfnamefont {Nikolay}\ \bibnamefont
  {Gromov}}, \bibinfo {author} {\bibfnamefont {Vladimir}\ \bibnamefont
  {Kazakov}}, \bibinfo {author} {\bibfnamefont {Sébastien}\ \bibnamefont
  {Leurent}}, \ and\ \bibinfo {author} {\bibfnamefont {Dmytro}\ \bibnamefont
  {Volin}},\ }\bibfield  {title} {\enquote {\bibinfo {title} {{Quantum spectral
  curve for arbitrary state/operator in AdS$_{5}$/CFT$_{4}$}},}\ }\href
  {\doibase 10.1007/JHEP09(2015)187} {\bibfield  {journal} {\bibinfo  {journal}
  {JHEP}\ }\textbf {\bibinfo {volume} {09}},\ \bibinfo {pages} {187} (\bibinfo
  {year} {2015})},\ \Eprint {http://arxiv.org/abs/1405.4857} {arXiv:1405.4857
  [hep-th]} \BibitemShut {NoStop}%
\bibitem [{\citenamefont {Leigh}\ and\ \citenamefont
  {Strassler}(1995)}]{Leigh:1995ep}%
  \BibitemOpen
  \bibfield  {author} {\bibinfo {author} {\bibfnamefont {Robert~G.}\
  \bibnamefont {Leigh}}\ and\ \bibinfo {author} {\bibfnamefont {Matthew~J.}\
  \bibnamefont {Strassler}},\ }\bibfield  {title} {\enquote {\bibinfo {title}
  {{Exactly marginal operators and duality in four-dimensional N=1
  supersymmetric gauge theory}},}\ }\href {\doibase
  10.1016/0550-3213(95)00261-P} {\bibfield  {journal} {\bibinfo  {journal}
  {Nucl. Phys.}\ }\textbf {\bibinfo {volume} {B447}},\ \bibinfo {pages}
  {95--136} (\bibinfo {year} {1995})},\ \Eprint
  {http://arxiv.org/abs/hep-th/9503121} {arXiv:hep-th/9503121 [hep-th]}
  \BibitemShut {NoStop}%
\bibitem [{\citenamefont {Frolov}(2005)}]{Frolov:2005dj}%
  \BibitemOpen
  \bibfield  {author} {\bibinfo {author} {\bibfnamefont {Sergey}\ \bibnamefont
  {Frolov}},\ }\bibfield  {title} {\enquote {\bibinfo {title} {{Lax pair for
  strings in Lunin-Maldacena background}},}\ }\href {\doibase
  10.1088/1126-6708/2005/05/069} {\bibfield  {journal} {\bibinfo  {journal}
  {JHEP}\ }\textbf {\bibinfo {volume} {05}},\ \bibinfo {pages} {069} (\bibinfo
  {year} {2005})},\ \Eprint {http://arxiv.org/abs/hep-th/0503201}
  {arXiv:hep-th/0503201 [hep-th]} \BibitemShut {NoStop}%
\bibitem [{\citenamefont {Beisert}\ and\ \citenamefont
  {Roiban}(2005)}]{Beisert:2005if}%
  \BibitemOpen
  \bibfield  {author} {\bibinfo {author} {\bibfnamefont {N.}~\bibnamefont
  {Beisert}}\ and\ \bibinfo {author} {\bibfnamefont {R.}~\bibnamefont
  {Roiban}},\ }\bibfield  {title} {\enquote {\bibinfo {title} {{Beauty and the
  twist: The Bethe ansatz for twisted N=4 SYM}},}\ }\href {\doibase
  10.1088/1126-6708/2005/08/039} {\bibfield  {journal} {\bibinfo  {journal}
  {JHEP}\ }\textbf {\bibinfo {volume} {08}},\ \bibinfo {pages} {039} (\bibinfo
  {year} {2005})},\ \Eprint {http://arxiv.org/abs/hep-th/0505187}
  {arXiv:hep-th/0505187 [hep-th]} \BibitemShut {NoStop}%
\bibitem [{\citenamefont {Kazakov}\ \emph {et~al.}(2015)\citenamefont
  {Kazakov}, \citenamefont {Leurent},\ and\ \citenamefont
  {Volin}}]{Kazakov:2015efa}%
  \BibitemOpen
  \bibfield  {author} {\bibinfo {author} {\bibfnamefont {Vladimir}\
  \bibnamefont {Kazakov}}, \bibinfo {author} {\bibfnamefont {Sebastien}\
  \bibnamefont {Leurent}}, \ and\ \bibinfo {author} {\bibfnamefont {Dmytro}\
  \bibnamefont {Volin}},\ }\bibfield  {title} {\enquote {\bibinfo {title}
  {{T-system on T-hook: Grassmannian Solution and Twisted Quantum Spectral
  Curve}},}\ }\href@noop {} {\  (\bibinfo {year} {2015})},\ \Eprint
  {http://arxiv.org/abs/1510.02100} {arXiv:1510.02100 [hep-th]} \BibitemShut
  {NoStop}%
\bibitem [{\citenamefont {Gromov}\ and\ \citenamefont
  {Levkovich-Maslyuk}(2015)}]{Gromov:2015dfa}%
  \BibitemOpen
  \bibfield  {author} {\bibinfo {author} {\bibfnamefont {Nikolay}\ \bibnamefont
  {Gromov}}\ and\ \bibinfo {author} {\bibfnamefont {Fedor}\ \bibnamefont
  {Levkovich-Maslyuk}},\ }\bibfield  {title} {\enquote {\bibinfo {title}
  {{Quantum Spectral Curve for a Cusped Wilson Line in N=4 SYM}},}\ }\href@noop
  {} {\  (\bibinfo {year} {2015})},\ \Eprint {http://arxiv.org/abs/1510.02098}
  {arXiv:1510.02098 [hep-th]} \BibitemShut {NoStop}%
\bibitem [{Note1()}]{Note1}%
  \BibitemOpen
  \bibinfo {note} {A large \(q\) expansion was considered in \protect \cite
  {Gromov:2010dy} for \(\beta \)-deformation}\BibitemShut {NoStop}%
\bibitem [{\citenamefont {Tseytlin}\ and\ \citenamefont
  {Zarembo}(1999)}]{Tseytlin:1999ii}%
  \BibitemOpen
  \bibfield  {author} {\bibinfo {author} {\bibfnamefont {Arkady~A.}\
  \bibnamefont {Tseytlin}}\ and\ \bibinfo {author} {\bibfnamefont
  {K.}~\bibnamefont {Zarembo}},\ }\bibfield  {title} {\enquote {\bibinfo
  {title} {{Effective potential in nonsupersymmetric SU(N) x SU(N) gauge theory
  and interactions of type 0 D3-branes}},}\ }\href {\doibase
  10.1016/S0370-2693(99)00471-2} {\bibfield  {journal} {\bibinfo  {journal}
  {Phys. Lett.}\ }\textbf {\bibinfo {volume} {B457}},\ \bibinfo {pages}
  {77--86} (\bibinfo {year} {1999})},\ \Eprint
  {http://arxiv.org/abs/hep-th/9902095} {arXiv:hep-th/9902095 [hep-th]}
  \BibitemShut {NoStop}%
\bibitem [{\citenamefont {Dymarsky}\ \emph {et~al.}(2005)\citenamefont
  {Dymarsky}, \citenamefont {Klebanov},\ and\ \citenamefont
  {Roiban}}]{Dymarsky:2005uh}%
  \BibitemOpen
  \bibfield  {author} {\bibinfo {author} {\bibfnamefont {A.}~\bibnamefont
  {Dymarsky}}, \bibinfo {author} {\bibfnamefont {I.~R.}\ \bibnamefont
  {Klebanov}}, \ and\ \bibinfo {author} {\bibfnamefont {R.}~\bibnamefont
  {Roiban}},\ }\bibfield  {title} {\enquote {\bibinfo {title} {{Perturbative
  search for fixed lines in large N gauge theories}},}\ }\href {\doibase
  10.1088/1126-6708/2005/08/011} {\bibfield  {journal} {\bibinfo  {journal}
  {JHEP}\ }\textbf {\bibinfo {volume} {08}},\ \bibinfo {pages} {011} (\bibinfo
  {year} {2005})},\ \Eprint {http://arxiv.org/abs/hep-th/0505099}
  {arXiv:hep-th/0505099 [hep-th]} \BibitemShut {NoStop}%
\bibitem [{\citenamefont {Fokken}\ \emph
  {et~al.}(2014{\natexlab{a}})\citenamefont {Fokken}, \citenamefont {Sieg},\
  and\ \citenamefont {Wilhelm}}]{Fokken:2013aea}%
  \BibitemOpen
  \bibfield  {author} {\bibinfo {author} {\bibfnamefont {Jan}\ \bibnamefont
  {Fokken}}, \bibinfo {author} {\bibfnamefont {Christoph}\ \bibnamefont
  {Sieg}}, \ and\ \bibinfo {author} {\bibfnamefont {Matthias}\ \bibnamefont
  {Wilhelm}},\ }\bibfield  {title} {\enquote {\bibinfo {title}
  {{Non-conformality of ${{\gamma }_{i}}$-deformed N = 4 SYM theory}},}\ }\href
  {\doibase 10.1088/1751-8113/47/45/455401} {\bibfield  {journal} {\bibinfo
  {journal} {J. Phys.}\ }\textbf {\bibinfo {volume} {A47}},\ \bibinfo {pages}
  {455401} (\bibinfo {year} {2014}{\natexlab{a}})},\ \Eprint
  {http://arxiv.org/abs/1308.4420} {arXiv:1308.4420 [hep-th]} \BibitemShut
  {NoStop}%
\bibitem [{\citenamefont {Jin}(2013)}]{Jin:2013baa}%
  \BibitemOpen
  \bibfield  {author} {\bibinfo {author} {\bibfnamefont {Qingjun}\ \bibnamefont
  {Jin}},\ }\bibfield  {title} {\enquote {\bibinfo {title} {{The Emergence of
  Supersymmetry in $\gamma_i$-deformed ${\cal N}=4$ super-Yang-Mills
  theory}},}\ }\href@noop {} {\  (\bibinfo {year} {2013})},\ \Eprint
  {http://arxiv.org/abs/1311.7391} {arXiv:1311.7391 [hep-th]} \BibitemShut
  {NoStop}%
\bibitem [{\citenamefont {Pomoni}\ and\ \citenamefont
  {Rastelli}(2009)}]{Pomoni:2008de}%
  \BibitemOpen
  \bibfield  {author} {\bibinfo {author} {\bibfnamefont {Elli}\ \bibnamefont
  {Pomoni}}\ and\ \bibinfo {author} {\bibfnamefont {Leonardo}\ \bibnamefont
  {Rastelli}},\ }\bibfield  {title} {\enquote {\bibinfo {title} {{Large N Field
  Theory and AdS Tachyons}},}\ }\href {\doibase 10.1088/1126-6708/2009/04/020}
  {\bibfield  {journal} {\bibinfo  {journal} {JHEP}\ }\textbf {\bibinfo
  {volume} {04}},\ \bibinfo {pages} {020} (\bibinfo {year} {2009})},\ \Eprint
  {http://arxiv.org/abs/0805.2261} {arXiv:0805.2261 [hep-th]} \BibitemShut
  {NoStop}%
\bibitem [{\citenamefont {Sieg}\ and\ \citenamefont
  {Wilhelm}(2016)}]{Sieg:2016vap}%
  \BibitemOpen
  \bibfield  {author} {\bibinfo {author} {\bibfnamefont {Christoph}\
  \bibnamefont {Sieg}}\ and\ \bibinfo {author} {\bibfnamefont {Matthias}\
  \bibnamefont {Wilhelm}},\ }\bibfield  {title} {\enquote {\bibinfo {title}
  {{On a CFT limit of planar $\gamma_i$-deformed $\mathcal{N}=4$ SYM
  theory}},}\ }\href {\doibase 10.1016/j.physletb.2016.03.004} {\bibfield
  {journal} {\bibinfo  {journal} {Phys. Lett.}\ }\textbf {\bibinfo {volume}
  {B756}},\ \bibinfo {pages} {118--120} (\bibinfo {year} {2016})},\ \Eprint
  {http://arxiv.org/abs/1602.05817} {arXiv:1602.05817 [hep-th]} \BibitemShut
  {NoStop}%
\bibitem [{Note2()}]{Note2}%
  \BibitemOpen
  \bibinfo {note} {The idea that our limit can be dominated by ``wheel'' graphs
  arose in common discussion with M.~Wilhelm and Ch.Sieg}\BibitemShut {NoStop}%
\bibitem [{\citenamefont {Broadhurst}(1985)}]{Broadhurst:1985vq}%
  \BibitemOpen
  \bibfield  {author} {\bibinfo {author} {\bibfnamefont {David~J.}\
  \bibnamefont {Broadhurst}},\ }\bibfield  {title} {\enquote {\bibinfo {title}
  {{Evaluation of a Class of Feynman Diagrams for All Numbers of Loops and
  Dimensions}},}\ }\href {\doibase 10.1016/0370-2693(85)90340-5} {\bibfield
  {journal} {\bibinfo  {journal} {Phys. Lett.}\ }\textbf {\bibinfo {volume}
  {B164}},\ \bibinfo {pages} {356} (\bibinfo {year} {1985})}\BibitemShut
  {NoStop}%
\bibitem [{\citenamefont {Fokken}\ \emph
  {et~al.}(2014{\natexlab{b}})\citenamefont {Fokken}, \citenamefont {Sieg},\
  and\ \citenamefont {Wilhelm}}]{Fokken:2014soa}%
  \BibitemOpen
  \bibfield  {author} {\bibinfo {author} {\bibfnamefont {Jan}\ \bibnamefont
  {Fokken}}, \bibinfo {author} {\bibfnamefont {Christoph}\ \bibnamefont
  {Sieg}}, \ and\ \bibinfo {author} {\bibfnamefont {Matthias}\ \bibnamefont
  {Wilhelm}},\ }\bibfield  {title} {\enquote {\bibinfo {title} {{A piece of
  cake: the ground-state energies in $\gamma_{i}$ -deformed $ \mathcal{N} $ = 4
  SYM theory at leading wrapping order}},}\ }\href {\doibase
  10.1007/JHEP09(2014)078} {\bibfield  {journal} {\bibinfo  {journal} {JHEP}\
  }\textbf {\bibinfo {volume} {09}},\ \bibinfo {pages} {78} (\bibinfo {year}
  {2014}{\natexlab{b}})},\ \Eprint {http://arxiv.org/abs/1405.6712}
  {arXiv:1405.6712 [hep-th]} \BibitemShut {NoStop}%
\bibitem [{\citenamefont {Ahn}\ \emph {et~al.}(2011)\citenamefont {Ahn},
  \citenamefont {Bajnok}, \citenamefont {Bombardelli},\ and\ \citenamefont
  {Nepomechie}}]{Ahn:2011xq}%
  \BibitemOpen
  \bibfield  {author} {\bibinfo {author} {\bibfnamefont {Changrim}\
  \bibnamefont {Ahn}}, \bibinfo {author} {\bibfnamefont {Zoltan}\ \bibnamefont
  {Bajnok}}, \bibinfo {author} {\bibfnamefont {Diego}\ \bibnamefont
  {Bombardelli}}, \ and\ \bibinfo {author} {\bibfnamefont {Rafael~I.}\
  \bibnamefont {Nepomechie}},\ }\bibfield  {title} {\enquote {\bibinfo {title}
  {{TBA, NLO Luscher correction, and double wrapping in twisted AdS/CFT}},}\
  }\href {\doibase 10.1007/JHEP12(2011)059} {\bibfield  {journal} {\bibinfo
  {journal} {JHEP}\ }\textbf {\bibinfo {volume} {12}},\ \bibinfo {pages} {059}
  (\bibinfo {year} {2011})},\ \Eprint {http://arxiv.org/abs/1108.4914}
  {arXiv:1108.4914 [hep-th]} \BibitemShut {NoStop}%
\bibitem [{Note3()}]{Note3}%
  \BibitemOpen
  \bibinfo {note} {We thank K.~Zarembo for the discussion and very helpful
  comments on this issue}\BibitemShut {NoStop}%
\bibitem [{\citenamefont {Panzer}(2015)}]{Panzer:2015ida}%
  \BibitemOpen
  \bibfield  {author} {\bibinfo {author} {\bibfnamefont {Erik}\ \bibnamefont
  {Panzer}},\ }\emph {\bibinfo {title} {{Feynman integrals and
  hyperlogarithms}}},\ \href
  {https://inspirehep.net/record/1377774/files/arXiv:1506.07243.pdf} {Ph.D.
  thesis},\ \bibinfo  {school} {Humboldt U., Berlin, Inst. Math.} (\bibinfo
  {year} {2015}),\ \Eprint {http://arxiv.org/abs/1506.07243} {arXiv:1506.07243
  [math-ph]} \BibitemShut {NoStop}%
\bibitem [{Note4()}]{Note4}%
  \BibitemOpen
  \bibinfo {note} {We are grateful to E. Panzer for his comments on this class
  of integrals.}\BibitemShut {Stop}%
\bibitem [{\citenamefont {Gromov}\ \emph {et~al.}(2009)\citenamefont {Gromov},
  \citenamefont {Kazakov},\ and\ \citenamefont {Vieira}}]{Gromov:2009tv}%
  \BibitemOpen
  \bibfield  {author} {\bibinfo {author} {\bibfnamefont {Nikolay}\ \bibnamefont
  {Gromov}}, \bibinfo {author} {\bibfnamefont {Vladimir}\ \bibnamefont
  {Kazakov}}, \ and\ \bibinfo {author} {\bibfnamefont {Pedro}\ \bibnamefont
  {Vieira}},\ }\bibfield  {title} {\enquote {\bibinfo {title} {{Exact Spectrum
  of Anomalous Dimensions of Planar N=4 Supersymmetric Yang-Mills Theory}},}\
  }\href {\doibase 10.1103/PhysRevLett.103.131601} {\bibfield  {journal}
  {\bibinfo  {journal} {Phys. Rev. Lett.}\ }\textbf {\bibinfo {volume} {103}},\
  \bibinfo {pages} {131601} (\bibinfo {year} {2009})},\ \Eprint
  {http://arxiv.org/abs/0901.3753} {arXiv:0901.3753 [hep-th]} \BibitemShut
  {NoStop}%
\bibitem [{\citenamefont {Bombardelli}\ \emph {et~al.}(2009)\citenamefont
  {Bombardelli}, \citenamefont {Fioravanti},\ and\ \citenamefont
  {Tateo}}]{Bombardelli:2009ns}%
  \BibitemOpen
  \bibfield  {author} {\bibinfo {author} {\bibfnamefont {Diego}\ \bibnamefont
  {Bombardelli}}, \bibinfo {author} {\bibfnamefont {Davide}\ \bibnamefont
  {Fioravanti}}, \ and\ \bibinfo {author} {\bibfnamefont {Roberto}\
  \bibnamefont {Tateo}},\ }\bibfield  {title} {\enquote {\bibinfo {title}
  {{Thermodynamic Bethe Ansatz for planar AdS/CFT: A Proposal}},}\ }\href
  {\doibase 10.1088/1751-8113/42/37/375401} {\bibfield  {journal} {\bibinfo
  {journal} {J. Phys.}\ }\textbf {\bibinfo {volume} {A42}},\ \bibinfo {pages}
  {375401} (\bibinfo {year} {2009})},\ \Eprint {http://arxiv.org/abs/0902.3930}
  {arXiv:0902.3930 [hep-th]} \BibitemShut {NoStop}%
\bibitem [{\citenamefont {Gromov}\ \emph {et~al.}(2010)\citenamefont {Gromov},
  \citenamefont {Kazakov}, \citenamefont {Kozak},\ and\ \citenamefont
  {Vieira}}]{Gromov:2009bc}%
  \BibitemOpen
  \bibfield  {author} {\bibinfo {author} {\bibfnamefont {Nikolay}\ \bibnamefont
  {Gromov}}, \bibinfo {author} {\bibfnamefont {Vladimir}\ \bibnamefont
  {Kazakov}}, \bibinfo {author} {\bibfnamefont {Andrii}\ \bibnamefont {Kozak}},
  \ and\ \bibinfo {author} {\bibfnamefont {Pedro}\ \bibnamefont {Vieira}},\
  }\bibfield  {title} {\enquote {\bibinfo {title} {{Exact Spectrum of Anomalous
  Dimensions of Planar N = 4 Supersymmetric Yang-Mills Theory: TBA and excited
  states}},}\ }\href {\doibase 10.1007/s11005-010-0374-8} {\bibfield  {journal}
  {\bibinfo  {journal} {Lett. Math. Phys.}\ }\textbf {\bibinfo {volume} {91}},\
  \bibinfo {pages} {265--287} (\bibinfo {year} {2010})},\ \Eprint
  {http://arxiv.org/abs/0902.4458} {arXiv:0902.4458 [hep-th]} \BibitemShut
  {NoStop}%
\bibitem [{\citenamefont {Arutyunov}\ and\ \citenamefont
  {Frolov}(2009)}]{Arutyunov:2009ur}%
  \BibitemOpen
  \bibfield  {author} {\bibinfo {author} {\bibfnamefont {Gleb}\ \bibnamefont
  {Arutyunov}}\ and\ \bibinfo {author} {\bibfnamefont {Sergey}\ \bibnamefont
  {Frolov}},\ }\bibfield  {title} {\enquote {\bibinfo {title} {{Thermodynamic
  Bethe Ansatz for the AdS(5) x S(5) Mirror Model}},}\ }\href {\doibase
  10.1088/1126-6708/2009/05/068} {\bibfield  {journal} {\bibinfo  {journal}
  {JHEP}\ }\textbf {\bibinfo {volume} {05}},\ \bibinfo {pages} {068} (\bibinfo
  {year} {2009})},\ \Eprint {http://arxiv.org/abs/0903.0141} {arXiv:0903.0141
  [hep-th]} \BibitemShut {NoStop}%
\bibitem [{Note5()}]{Note5}%
  \BibitemOpen
  \bibinfo {note} {The singularity at \(L=2\) in this formula signals the
  breakdown of the conformal invariance \protect \cite {Fokken:2013aea} but
  this \(L=2\) state does not appear in most of the multi-point correlators in
  planar limit.}\BibitemShut {Stop}%
\bibitem [{\citenamefont {Panzer}(2013)}]{Panzer:2013cha}%
  \BibitemOpen
  \bibfield  {author} {\bibinfo {author} {\bibfnamefont {Erik}\ \bibnamefont
  {Panzer}},\ }\bibfield  {title} {\enquote {\bibinfo {title} {{On the analytic
  computation of massless propagators in dimensional regularization}},}\ }\href
  {\doibase 10.1016/j.nuclphysb.2013.05.025} {\bibfield  {journal} {\bibinfo
  {journal} {Nucl. Phys.}\ }\textbf {\bibinfo {volume} {B874}},\ \bibinfo
  {pages} {567--593} (\bibinfo {year} {2013})},\ \Eprint
  {http://arxiv.org/abs/1305.2161} {arXiv:1305.2161 [hep-th]} \BibitemShut
  {NoStop}%
\bibitem [{Note6()}]{Note6}%
  \BibitemOpen
  \bibinfo {note} {We are grateful to K.~Zarembo for sharing with us his
  computations of single wrapping case using similar techniques}\BibitemShut
  {NoStop}%
\bibitem [{\citenamefont {Zamolodchikov}(1980)}]{Zamolodchikov1980a}%
  \BibitemOpen
  \bibfield  {author} {\bibinfo {author} {\bibfnamefont {A.~B.}\ \bibnamefont
  {Zamolodchikov}},\ }\bibfield  {title} {\enquote {\bibinfo {title}
  {{``Fishing-net'' diagrams as a completely integrable system}},}\ }\href
  {\doibase 10.1016/0370-2693(80)90547-X} {\bibfield  {journal} {\bibinfo
  {journal} {Phys. Lett.}\ }\textbf {\bibinfo {volume} {B97}},\ \bibinfo
  {pages} {63--66} (\bibinfo {year} {1980})}\BibitemShut {NoStop}%
\bibitem [{\citenamefont {Kostov}\ and\ \citenamefont
  {Staudacher}(1997)}]{Kostov:1996bs}%
  \BibitemOpen
  \bibfield  {author} {\bibinfo {author} {\bibfnamefont {Ivan~K.}\ \bibnamefont
  {Kostov}}\ and\ \bibinfo {author} {\bibfnamefont {Matthias}\ \bibnamefont
  {Staudacher}},\ }\bibfield  {title} {\enquote {\bibinfo {title}
  {{Two-dimensional chiral matrix models and string theories}},}\ }\href
  {\doibase 10.1016/S0370-2693(96)01664-4} {\bibfield  {journal} {\bibinfo
  {journal} {Phys. Lett.}\ }\textbf {\bibinfo {volume} {B394}},\ \bibinfo
  {pages} {75--81} (\bibinfo {year} {1997})},\ \Eprint
  {http://arxiv.org/abs/hep-th/9611011} {arXiv:hep-th/9611011 [hep-th]}
  \BibitemShut {NoStop}%
\bibitem [{\citenamefont {J.~Caetano}()}]{CGKtbp}%
  \BibitemOpen
  \bibfield  {author} {\bibinfo {author} {\bibfnamefont {V.~Kazakov}\
  \bibnamefont {J.~Caetano}, \bibfnamefont {O.~Gurdogan}},\ }\bibfield  {title}
  {\enquote {\bibinfo {title} {to be published},}\ }\href@noop {} {\
  }\BibitemShut {NoStop}%
\bibitem [{\citenamefont {Gromov}\ and\ \citenamefont
  {Levkovich-Maslyuk}(2011)}]{Gromov:2010dy}%
  \BibitemOpen
  \bibfield  {author} {\bibinfo {author} {\bibfnamefont {Nikolay}\ \bibnamefont
  {Gromov}}\ and\ \bibinfo {author} {\bibfnamefont {Fedor}\ \bibnamefont
  {Levkovich-Maslyuk}},\ }\bibfield  {title} {\enquote {\bibinfo {title}
  {{Y-system and $\beta$-deformed N=4 Super-Yang-Mills}},}\ }\href {\doibase
  10.1088/1751-8113/44/1/015402} {\bibfield  {journal} {\bibinfo  {journal} {J.
  Phys.}\ }\textbf {\bibinfo {volume} {A44}},\ \bibinfo {pages} {015402}
  (\bibinfo {year} {2011})},\ \Eprint {http://arxiv.org/abs/1006.5438}
  {arXiv:1006.5438 [hep-th]} \BibitemShut {NoStop}%
\end{thebibliography}%

\end{document}